# AI, Ageing and Brain-Work Productivity: Technological Change in Professional Japanese Chess

Eiji Yamamura[a] and Ryohei Hayashi[b]*

[a]Affiliation
[b]Affiliation

*Corresponding author:

Declaration of interest:

Funding:

# AI, Ageing and Brain-Work Productivity: Technological Change in Professional Japanese Chess

Summary: Using Japanese professional chess (Shogi) players' records in the novel setting, this paper examines how and the extent to which the emergence of technological changes influences the ageing and innate ability of players' winning probability. We gathered games of professional Shogi players from 1968 to 2019, which we divided into three periods: 1968–1989 (before the emergence of information technology), 1990–2012 (the diffusion of information and communication technology) and 2013–2019 (diffusion of artificial intelligence technology). Upon comparing these three periods, the major findings are: (1) diffusion of artificial intelligence (AI) reduces innate ability, which reduces the performance gap among same-age players; (2) in all the periods, players' winning rates declined consistently from 20 years and as they get older; (3) AI accelerated the ageing declination of the probability of winning, which increased the performance gap among different aged players; (4) the effects of AI on the ageing declination and the probability of winning are observed for high innate skill players but not for low innate skill ones. This implies that the diffusion of AI hastens players' retirement from active play, especially for those with high innate abilities. Thus, AI is a substitute for innate ability in brain-work productivity.

# 1. Introduction

How will new technological progress, such as information and communications technology (ICT) and artificial intelligence (AI), change the work environment and the labour market? Several studies have tried to answer this question (e.g., Autor, 2015; Autor et al. 2003; Autor & Dorn, 2013; Ikenaga & Kambayashi). The impact of technological progress varies according to jobs and required skills. For example, the impact of AI on board game players has been remarkable (Fujita, 2018; Hamaguchi & Kondo, 2018). In 1997, IBM's AI—named Deep Blue—beat Garry Kasparov, the world's chess champion. The rules of the Japanese chess (Shogi) game are more complicated than for traditional chess and the possible choices per game are far higher. Multiple professional Shogi players did not predict that they would be defeated by AI (Japan Shogi Federation (Nihon Shogi Renmei), 1996). Owing to the limitations in technology in 2007, AI could not win against Akira Watanabe, who was a top-class professional Shogi player and the major title 'Ryuo' holder. Based on his experience in the game, Watanabe predicted that AI could not win against professional Shogi players. However, the level of AI is almost equivalent to the top-notch amateur Shogi players (Hoki & Watanabe, 2007). After 6 years, in 2013, AI defeated professional Shogi player, Shin-ichi Sato. Since then until 2017, various professional Shogi players have played against AI but could not win in most cases. The results of the AI vs. Human games indicated that AI could surpass professional Shogi players. Akira Watanabe accepted that AI has surpassed the top-rank professional Shogi players and confessed that he became an active user of AI 14 years ago to improve his Shogi skills in games against AI (Watanabe, 2021).

Several studies have dealt with how technological progress influences productivity in the labour market (Acemoglu, Autor, et al. 2020; Acemoglu, Lelarge, et al. 2020; Acemoglu & Autor, 2011; Acemoglu & Restrepo, 2018, 2020; Autor, 2015; Autor et al. 2003; Autor & Dorn, 2013). In Europe, polarisation trends in the labour market appeared from 1993 to 2010 because of an increase in high and low-paying occupations, whereas middle-paying occupations decreased (Autor, 2014). Similarly, non-routine tasks have increased in Japan, while routine tasks have decreased from 1960 to 2005. This suggests a long-term labour market polarisation (Ikenaga & Kambayashi, 2016). Technology leads to wage inequality and polarisation in the labour market (Acemoglu & Autor, 2011; Autor, 2015).

The recent development of AI technology allows it to replicate the human brain and, therefore, would replace brain work. An estimated at 47% of jobs are expected to be at the risk of computerisation (Frey & Osborne, 2017). Meanwhile, there is an argument that complementarity between labour and robots increases productivity (Autor, 2015).[1] The skills and technology required for work differ according to jobs, industries, and stages of economic development. Accordingly, the influence of technology diffusion varies according to the settings where workers are confronted. It is valuable to scrutinise how technology diffusion influences workers' performance and the inequality between workers with different skills within the labour market in a specific industry by comparing different periods.

---

[1] Most firms consider AI technology and robots to save on labour, whereas firms employing high-skilled workers expect the emergence of new jobs which do not exist presently (Morikawa, 2017).

In a professional board game, various factors that influence labour productivity can be controlled under the novel setting, especially in a two-player game such as chess, where a player's performance does not depend on a team mate's performance. During the game, players cannot receive the advice of others and use technological devices.[2] Innate ability, a trait or characteristic present at birth, is one of the key factors determining individual performance and, hence, labour productivity. Apart from it, human capital is accumulated through learning from experience, which improves productivity, whereas mental ability declines with age, which lowers productivity.[3] The relationship between age and performance is an inverted U-shape profile with a peak at approximately 21 years for chess players' productivity (Bertoni et al. 2015). Ageing players consider AI more difficult to use and are less likely to catch up with drastic environmental changes. However, it is unknown whether technological progress, such as ICT and AI, influences innate ability and ageing on player's performance.[4] We construct game-level data from 1968 to 2019 to compare ageing and innate ability effects between three sub-periods. From the data, the major findings are that performance polarisation is observed in the period when AI is widely used to improve strategies among professional Shogi players. The following mechanism explains this. AI is a substitute for the innate ability of an individual player, which reduces the gap in performance between players of the same age. However, the effects of negative ageing on performance are strengthened by the diffusion of AI, inducing players to retire from active play earlier than in other periods. These imply that the polarisation of players' performance depends on whether players can make the best use of AI rather than innate ability.

The remainder of this article is organised as follows. Section 2 overviews the real situation of professional Shogi. Section 3 explains the dataset and presents the basic statistics. Section 4 proposes testable hypotheses and describes the empirical method. Section 5 presents and interprets the estimated results. The final section offers some reflections and conclusions.

## 2. Overview of Professional Shogi

### 2.1. Rules and differences from international Chess

On some points, the rules of Shogi are the same as that of international chess. The game's goal is for one player to checkmate the other player's king, winning the game. Several Shogi pieces can be moved like those of international chess. However, Shogi is different from international chess as follows. In cases where a piece occupies a legal destination for an opposing piece, it may be captured by replacing it with the opposing piece. Captured pieces are retained in hand and can be brought back into play under the capturing player's control. On any turn, instead of moving a piece on the board, a player may select a piece in hand and place it on any empty square. Therefore, one of that player's active pieces on the board can be moved. This is called dropping the piece or simply, a drop. A drop counts as a complete move. The ability to drop in Shogi drastically increases the player's choices, which leads to more strategic variety and complexity than in international chess. Naturally, AI won against a professional Shogi player 16 years later than Deep Blue won against Kasparov in 1997.

---

[2] There are some exceptional cases. In chess, where various countries' players participate in the games, political ideology is influential (Frank et al. 2004; Frank & Krabel, 2013). In addition, players collide with the same-country players to improve country-level productivity (Moul & Nye, 2009). However, professional Shogi is not yet open to foreign players; therefore, political factors do not influence the games.
[3] Researchers examined how chess instruction changes children's educational attainment. In developed countries, the effect is not observed (Jerrim et al. 2018), whereas, in developing countries, the positive effect is observed to a certain extent (Islam et al. 2021).
[4] Peer effects from 'Superstar' were observed in international Chess. AI seems to reduce the effect (Bilen & Matros, 2021).

The difference in the rule is why the Shogi game rarely ends in a draw. In the professional Shogi games from 1968 to 2019, the rate of draw games was 0.7% which is far lower than the 53.4% draw rate for international Chess from 1970 to 2017.[5] The index of a player's performance is made by wins, losses, and draws, owing to the extremely high draw rates in previous international chess studies (Bertoni et al. 2015). However, as opposed to chess, Shogi players are unlikely to develop a strategy to draw. In this study, it is not necessary to consider draw rates because draws rarely occur in Shogi.

## 2.2. System

Shogi players are promoted to professional status from the 'Shoreikai League', where young amateur players are selectively qualified to enter. Regardless of gender, promising amateurs can apply for the 'Shoreikai'. Members of the Shoreikai are neither professional players nor amateur players because they all aspire to survive the fierce struggle for existence to become professional players. The Shoreikai consists of approximately 200 players. Players younger than 15 years qualify for an entrance examination held once a year in August. In the first stage, a candidate plays games with 6 other candidates. They proceed to the second stage if they gain a majority of wins. In the second stage, the candidate plays games with 3 incumbent Shoreikai members. The candidate passes the examination if the candidate wins at least one game. There are 30–40 candidates and the ratio of successful applicants is approximately 10%–20%.[6] In most cases, Shoreikai players start their career from the bottom grade 6-kyu directly after passing the entrance examination. There are 9 grades in Shoreikai, and the top grade is 3-dan.[7] Members can proceed to the next higher grade if they gain high winning rates in the league of each grade. Approximately 30 members constitute the top league, the 3-dan league. Among them, 2 members can become professional players in half a year only if they gain the first or the second position in the 3-dan league. After successfully being winners in the final stage of Shoreikai, they become 4-dan grade, which means they can enter the professional league. Therefore, every year, only four Shoreikai members can become professional players. Further, members must withdraw from the Shoreikai, regardless of their intention, at 26 years and cannot proceed to the professional level. The Shoreikai system forms a straight gate to enter the world of professional Shogi, even though most Shoreikai players could not survive and dropped out from the Shoreikai league before becoming professional players.

---

[5] Qiyu Zhou. Has the number of draws in chess increased? Chess News. https://en.chessbase.com/post/has-the-number-of-draws-in-chess-increased. (accessed on March 10, 2022).
[6] See reports of former Shoreikai members. https://www.i-tsu-tsu.co.jp/blog/shoureikai/ (accessed on March 10, 2022).
[7] The higher grade is, the grade number of 'kyu' reduces. That is, starting from 6-kyu, 5- kyu, 4-kyu, 3-kyu, 2-kyu, 1-kyu. After 1 kyu, the grade number of 'dan' increases. That is, 1-dan, 2-dan and 3-dan.

The number of professional players is kept constant. Hence, four professional players retire every year to be replaced by four players promoted from the Shoreikai. As a retirement rule, professional players are forced to retire if their winning rate has been extremely low for several years. However, it is not difficult to keep their professional status once they become professional players. There were 301 professional Shogi players from 1968 to 2019. Only 7 players have retired below 45 years, owing to their poor performance. Thus, most of them could play as professionals for 45 years even if their performances were low. The winning rate of professional players declines as they get older, and naturally, they retire in their 50s or 60s. Consequently, below 45, a professional players' perfect record can be obtained from their debuts. This is a setting unique to professional Shogi players, different from the international chess league, where teenagers can participate and drop out frequently (Bertoni et al., 2015). Therefore, members of the professional league hardly change before they turn 45 years old because they survive in the highly competitive Shoreikai league before becoming professional players.

Membership to the Shoreikai is open to male and female players. However, no female player has yet accomplished this feat. Thus, all professional players are males,[8] partly because of the very competitive environment in the Shoreikai and the gender difference in performance in a mixed-gender competitive environment (Booth & Yamamura, 2018). No female Shogi players can win through in Shoreikai and be promoted to become professional players. The situation is different from chess where male and female compete in games (Dilmaghani, 2020, 2021; Dreber et al., 2013; Dreber, Gerdes, Gränsmark, et al. 2013; Gerdes & Gränsmark, 2010; Gränsmark, 2012),[9]

After becoming professionals, Shogi players ordinary enter the league to play for the Meijin title, called 'Jun-i Sen'. In this league, there are 5 classes: 'C2 (bottom)', 'C1', 'B2', 'B1', 'A (top)'. Similar to Shoreikai, only the first and second ranks of winning rates in a class can be promoted to a higher class. The last and second to the last rank players automatically move to the lower class. The change of class occurs once a year and is considered to reflect the player's strength.[10] It at least takes 5 years to become a class 'A' member after entering the professional league. The number of class 'A' players is fixed at 10. The player with the highest winning rate in the season has the right to play with the title holder of Meijin. Therefore, a player cannot play with the title holder of Meijin if he does not belong to 'A' even if he is stronger than any 'A' class player. Apart from 'Meijin', there are 7 other titles, which include 'Ryuo', 'Oi', 'Oza', 'Kio', 'Eio', 'Osho' and 'Kisei'. Unlike Meijin, any professional player could get these titles if they survive the fierce struggle in the title tournament.[11] Besides the league and tournaments for the 8 major titles, there are different games for some minor titles.

---

[8] There is a separate and different system specially designed for female professionals. Therefore, some female Shoreikai players can also play the game as 'Female Professionals' who do not have the right to participate in the professional league (Jun-i Sen League).
[9] Gender differences in performance can be considered to come partly from a kind of innate ability.
[10] Besides their class in the league, the status of players can be captured by 'dan'. In the professional world, there are 6 player grades: 4, 5, 6, 7, 8 and 9-dan. This reflects the experience of the total number of winnings from their debut. Hence, players' grades increase as they gain experience in games. However, the grade decreases if their performance is very low. Therefore, the degree of dan is unlikely to reflect the players' strength.

## 2.3. Features of the three periods

Accurate data is available from 1968, therefore, we limited the data before 1968 to avoid measurement errors. The professional Shogi environment from 1968 to 2019 can be divided into three periods in terms of technological development.

First, 1968–1989 was characterised by human-led Shogi before technology emerged. This period is equivalent to the later Showa era in Japan. Shogi players trained themselves to learn from books, including the records of historically famous professional games before ICT emerged. Hence, players' strategies and skills developed slowly. Rather than improving skills and strategies, players' real-life experience is considered important to win in a game because the psychological tactics of the game effectively destroy the opponents' morale. For instance, most players feel mental stress before a game, decreasing their appetite. In contrast, Yasuharu Oyama, a multiple title holder for many years, would intentionally eat a lot in front of his opponent before a game to display his toughness. The opponent often fell victim to Oyama's tactics and could not fully demonstrate his ability.

Second, the period 1990–2012 is characterised by the information-oriented Shogi. ICT had developed during this period covering the early and middle of the Heisei period. Players can search for the latest games records by using a computer database. Further, players frequently met together in the real world and exchanged opinions and investigate new strategies. Naturally, new strategies developed frequently and diffused earlier among professional players than in the Showa era. Players' skills and strategies drastically improved, incapacitating the Oyama tactics style and prevented players from adopting the same tactics. During this period, Yoshiharu Habu, who held multiple titles consistently, argued that 'Shogi is a pure board game and so real life experience does not contribute to a player's performance at all'.

Third, the period 2013–2019 is characterised by the AI-dominant Shogi. As explained in the introduction, in 2013, a professional Shogi player was first defeated by AI. Within the next several years, various top-ranked players also lost games to AI. In 2017, there were two games of AI vs. Amahiko Sato, who held the title of 'Meijin'—the most valuable title. Amahiko Sato suffered complete defeats in both games, implying that AI exceeded professional players. According to the major multiple title holder, Akira Watanabe, 'Before diffusion of AI, it took around 10 years to establish a standard move, whereas it takes only a day now. This results in an unexpected situation. Any players can use AI, leading to the gap in Shogi skills between players to be narrowed' (Watanabe, 2021). Owing to AI's diffusion, top-rank players are less likely to derive advantage from their innate ability.

---

[11] Among 8 major titles, 'Meijin' and 'Ryuo' are equally the highest status because of the longest history of 'Meijin' and the highest prize money of 'Ryuo'.

## 3. Data

We collected game-level data from the Shogi database through the internet.[12] The data information includes the game's date and results, kinds of game (name of the tournament, class of the league, final and semi-final), player's information, such as name, birth date, age of his debut as a professional player, Elo rating which is the index for player's strength, class to which they belong and grade (dan). The games database covered records from the 1950s. However, in the early period of the database, some games had errors or insufficient information. From 1968, information on games was accumulated. In previous Chess studies, the Elo rating is generally used to measure individual player's strength (Gerdes & Gränsmark, 2010; Minondo, 2017; Simkin & Roychowdhury, 2015).

Figure 1 demonstrates the Elo rates based on the three different periods defined in section 2.3. Its form is similar to the normal distribution for the periods 1968–1989 and 1990–2012. However, from 2013 to 2019, we observe a peak at 1600 and a small bump slightly over 1800. From Figure 1, we infer that in the period of AI, the strength of players is polarised. In Figure 2, the sample is divided into the very recent period 2018 to 2019 and the period 1968 to 2017 because AI has been widely diffused and utilised by players since 2017 (Watanabe, 2021). Twin peaks are observed from 2018 to 2019, whereas standard distribution is observed from 1968 to 2017. Strength polarisation appeared probably because some players became active users of AI for their investigation of Shogi, whereas others did not or could not effectively use AI. It is crucial whether players could catch up with the drastically changed environment.

The Elo rating is considered to be different from performance. According to Bertoni et al. (2015), 'ELO cannot be considered as a measure of productivity at Chess, which depends on realised rather than unexpected wins and draws... Rather than a measure of productivity, ELO is a measure of relative ability at Chess at a given point in time: it predicts ex-ante how likely a player is to win when he plays against an opponent, but it does not measure winning intensity' (Bertoni et al. 2015, p.48). However, the Elo rating changes to reflect the results of games after becoming professional players. Therefore, it captures the innate ability and learning effect from the player's experience.

As explained in section 2, it is difficult to enter a professional Shogi league, unlike in international Chess. During the 1968–2019 period, all the titleholders of the Meijin debuted as professional players when they were teenagers, younger than the mean debut age of approximately 21 years, as shown in Table 1. Hence, the younger the player started, the higher their innate ability. Under the novel setting of the professional Shogi system, the innate ability can be captured by the debut ages of the professional players. Further, similar to international Chess, self-selection bias possibly occurs in professional Shogi (Bertoni et al., 2015). However, only 0.7% retired below 45 years in the studied period owing to their poor performance. Suppose we restrict the sample to professional players; the self-selection bias is unlikely to influence the results. This study uses a full sample covering aged players and a sub-sample limited to players younger than 45 years.

---

[12] Web address of Shogi database is http://kenyu1234.php.xdomain.jp/menu.php (Accessed in December 2019).

Figure 3 illustrates how total winning rates change according to players' debut years. Using the full sample, we observe that the average debut year's winning rate is approximately 0.7, implying that the rookie professional player's winning rate is approximately 70%. Players' winning rate reduced to approximately 50% when they debuted at 22 years and further declined slightly below 40% when they debuted at 26 years. Winning rates tend to decline as debut age increases, using the full and the sub-samples below 45 years. There is no statistically significant difference between the samples. The opponent's strength changes according to age because players with higher performance are likely to play with similar level players to compete for a major title in a higher class league.

Figure 4 demonstrates that the winning rate consistently declines as players become older. The winning rate based on the sub-sample below 45 is lower than that of the full sample. In the sub-sample, all players are below 45 years; therefore, games between players over 45 and below 45 are not included. Game records where younger and older players played with lower winning rates are excluded. Further, the self-selection effect hardly occurs when the sub-sample is used, reducing upward bias. In a previous study, the opponent's strength is controlled using Elo ratings (Bertoni et al. 2015). In the Appendix, Figure A1 is illustrated by calculating the index. In Figure A1, we observed a similar trend after controlling the opponent's Elo rating, consistent with Bertoni et al. (2015).

Figures 5 and 6 use the full sample to compare the three periods. Figure 5 illustrates debut ages and winning rates relation. A decline in the total winning rate was observed in 1968–1989 and 1990–2012. However, in the period 2013–2019, the gap in total winning rate between debut ages is smaller than in the other periods, although the winning rate slightly declines as debut age gets older. Therefore, the impact of the innate ability on winning rate is smaller in the period when skill and strategy improved drastically. The diffusion of AI reduces the advantage of innate ability because players with lower innate abilities utilise AI to catch up with those with higher innate abilities. AI can be used to reduce the gap in innate abilities between players. Figure 6 illustrates the ages and the winning rates relationship. The winning rate constantly declined during the 1968–1989 period as players got older. However, winning rates hardly changed for players older than 50 years. There are two possible reasons why the winning rates remained constant at an advanced age. First, players learn the psychological tactics of the game through their experiences, which effectively maintains their winning rates. Second, more able players can survive despite their advanced ages, whereas other aged players are forced to retire owing to poor performance. During the 1990–2012 and 2013–2019 periods, the decline in the total winning rate is constantly observed. Even in these periods, the self-selection effect existed but did not maintain the aged player's winning rate. Overall, the three periods jointly suggest that drastic improvement in skills and strategies owing to ICT or AI reduces the effect of learning psychological tactics, resulting in a consistent decline in the winning rates.

## 4. Hypothesis and Method

## 4.1. Hypotheses

In Figures 1–6, we can observe that players' performance changes. From these observations and the environmental change of professional Shogi, we proposed several testable hypotheses in this section. Following Figure 1 and the findings from international Chess (Bertoni et al. 2015), mental productivity of board games declines approximately after 20 years; we proposed *Hypothesis 1*.

*Hypothesis 1. Winning rates decline as players get older after their debut.*

The diffusion of AI distinctly accelerates skills and strategy development. Therefore, older players face a challenge catching up with environmental change after the diffusion of AI. We proposed *Hypothesis 2*.

*Hypothesis 2. Declining of winning rates due to ageing is more rapid after AI diffusion than ever before.*

## 4.2. Method

In the model, the estimated function takes the following form to examine how ages and innate ability influence players' performance.

$$WIN_{igt} = \alpha_0 + \alpha_1 AGE_{igt} + \alpha_2 AGE^2_{igt} + \alpha_3 AGE\_OP_{igt} + \alpha_4 AGE\_OP^2_{igt}$$
$$+ \alpha_5 DEB\_AGE_i + \alpha_6 DED\_AGE^2_i + \alpha_7 DEB\_AGE\_OP_i + \alpha_8 DED\_AGE\_OP_i^2 + \alpha_9 ELO_{igt} + \alpha_9 ELO\_OP_{igt} + X'_{it}B + u_i.$$

In the function, the suffix '*i*' is an individual player, suffix '*g*' is a game, and suffix '*t*' is the time point. Elo rating can be a player's ability that differs from performance (Bertoni et al. 2015). WIN is the first player's winning dummy which is 1 if the first player wins; otherwise, 0. Table 1 indicates that the mean value of WIN (win dummy) is 0.53, implying that the first player's winning rate is 53%, while the second player's is 47%. Therefore, the first players enjoy some advantage, although the difference between the first and second players is small. In professional Shogi, it is randomly determined whether a player is first or second. Hence, as shown in Table 1, the first player's age in the game and his debut age are similar to those of the second player.

As explained in section 2, unlike international Chess, the draw rate is only 0.7%, so draw games are not included in the sample. Compared to international Chess, a player's productivity can be captured by a player's win because it is unnecessary to consider a draw. However, it is critical to control the opponent's and player's Elo ratings. Furthermore, the player's productivity is calculated by considering the opponent's Elo rating (Bertoni et al. 2015, p.48). As seen in the function, this study includes both Elo ratings as independent variables. Further, it is more convenient to interpret the results when the probability of winning is investigated. Therefore, we use the winning dummy for a game (WIN) as a proxy variable for an individual's performance and dependent variable in the model. The Probit model is used for this estimation because the dependent variable is 0 or 1.

This study examines the relationship between players' ages and their performances. Shogi is a two-player game; we should consider both players' characteristics. The opponent (second player) with lower ability increases the probability of the first player. Therefore, the predicted sign of the first player's characteristics is opposed to that of the second player. The ages of the first player (AGE) and that of the second player (AGE_OP) are included as key variables. From *Hypothesis 1*, the expected signs of AGE and AGE_OP are negative and positive, respectively. To capture first the players' innate ability and that of his opponent, the debut age of the first (DEB_AGE) and second player (DEB_AGE_OP) are included. The older the debut age, the lower the innate ability. Hence, the expected coefficient's sign is negative and positive for DEB_AGE and DEB_AGE_OP, respectively. The relationship between the WIN and these variables is potentially non-linear. For this reason, we incorporated their squared terms such as $DEB\_AGE^2$, $DEB\_AGE\_OP^2$, $AGE^2$, and $AGE\_OP^2$. However, for convenience of interpretation, their results of marginal effect are divided by 100. Therefore, ELO and ELO_OP are incorporated, and their expected sign is positive and negative, respectively.

$X_i$ represents the control variable vector, and B is the vector of their coefficients. X consists of the time point of the game, dummies for the first and the second players' class of the league, dummies for the first and the second players' 'dan'. These variables also control for players' strength. Besides it, we add dummies for the status and classification of games because the importance of games widely varies. For instance, it is much more important to win a championship game of a major title than a preliminary game.

Further, based on the Probit model results in each period, we provide simulation analysis to visually present how a player's performance differs regarding his innate ability and how a player's performance changes as he gets older. We then compare visualised changes in performance between three periods to consider how the path of a player's performance declination differs according to technologies such as ICT and AI.

## 5. Results
## 5.1. Results of the Probit model

Tables 2–4 show that the Probit model results and that the values without parentheses in each variable are marginal effects. Table 2 shows the results based on the sample covering all ages. The results of Table 3 are based on the sub-sample restricting players who are 45 years or below to mitigate the self-selection biases. Further, the sub-sample is divided into players who debuted at 20 years or below and players who debuted above 20 years. Then, in Table 4, Panel A presents results using the sub-sample of players who debuted at an early age, considered to have high innate ability. Panel B indicates results using the sub-sample of players who debuted at a later age, considered to have low innate ability.

Firstly, from Table 2, we found that the results of most variables show the expected sign. Furthermore, in most variables, statistical significance is observed in columns (1) and (2), whereas statistical significance is not observed in many of the variables in column (3). Looking closely at column (3) tells that AGE and AGE_OP are statistically significant, whereas their squares are not. This implies that the relationship between ages and the winning rate is negative and linear in the 2013–2019 period, while the relationship is negative and non-linear before the diffusion of AI. The marginal effect of AGE is −0.022 or −0.023, implying that the first player's probability of winning is rescued by approximately 2.2% if he gets a year older. However, such reduction in winning rate is mitigated as players become older in the periods before AI diffusion because $AGE^2$ shows a significant positive sign. Regarding DEB_AGE and its square ($DEB\_AGE^2$), these show significant expected signs in columns (1) and (2) but not in column (3). Meanwhile, DEB_AGE_OP and its square ($DEB\_AGE\_OP^2$) indicate the predicted significant sign in columns (3) but not in columns (1) and (2). Consistent with this prediction, ELO and ELO_OP show the expected positive and negative signs, respectively, which are statistically significant at the 1% level in all columns.

In Table 3, the variable of player's age shows the predicted sign and statistical significance, except for $AGE,^2$ in column (3). The variable of debut ages show the expected signs, although not statistically significant for some of them in columns (1) and (2). The results for the variable of Elo rating is similar to that in Table 2. Even after dividing the sample into high and low ability players, as shown in Table 4, the results are similar to those in Table 3.

## 5.2. Simulation

As for the simulation about the relationship between the winning rate and debut age, the path is illustrated from 16 to 26 years. From the results in Table 2, we calculate the predicted values of winning rates and visualise the effects of the debut age on the winning rates in Figure 7. Notably, for a base of values, we use the mean values of the winning rate of those who debuted at 21 years because a professional player's mean debut age is 21. The winning rates at the age of 21 differ according to the periods, 0.59 (1968–1989), 0.63 (1990–2012) and 0.63 (2013–2019). The winning rates of players who debuted earlier at 16 or 17 years are thought to suffer upward bias because they are more able than most players who had not yet become professional before 20 years. Then, statistically significant marginal effects of the debut ages are used for simulation.[13] Figure 8 is illustrated using the sub-sample equal to or less than 45 years. Accordingly, in

---

[13] Let us analyse the related part of estimated function, $\alpha_5$ DEB_AGE $_i$ + $\alpha_6$ DED_AGE$^2$ $_i$ + $\alpha_7$ DEB_AGE_OP $_i$ + $\alpha_8$ DED_AGE_OP $_i$$^2$. If all of these results are statistically significant, we calculate the first player's winning rate at 26 years old as;
'The mean winning rate at 21 years +{ $\alpha_5×5+\alpha_6×5^2−(\alpha_7×5+\alpha_8×5^2)$}/2'.
Age 26 is 5 years older than that of 21, and so its difference is used for the prediction. Hence, for calculation at age 16, we put '−5'. Further, the second player's debut age effects are opposite to that of the first players. The probability of being the first ( second) player is 0.5. Therefore, the marginal effect of the first (second) player's debut age is weighted by 0.5. Hence, we made the sign of second player's variables opposite. Hence, in the case of the period 1968–1989, the winning rate at 26 years is obtained by;

Figures 7 and 8, the relationship between the debut age and the winning rates varies according to the periods. However, the slope in the period of AI diffusion is less steep than in the other periods, consistent with Figure 5. In Figure 8, interestingly, the slope of the AI diffusion period is almost equal to that of the period in the Showa low-technology period. However, the winning rate in the AI period is higher than in the low-technology period, if we compare the winning rates of the same ages. Meanwhile, the winning rate at 16 years is remarkably higher, and the slope is steeper in the ICT diffusion period than in other periods. Hence, compared to the AI diffusion period, the winning rate gap is remarkably larger in the period directly before AI emergence. This implies that AI contributes to reducing the performance gap between high and low innate ability players, whereas the ICT increased the gap.

Let us now consider the relationship between the winning rate and the players' ages. The initial value of the winning rate is the mean value of debut ages at 21 when the players averagely debuted. The marginal effects of ages that are statistically significant are used for simulation. The initial values of the winning rates and the marginal effects differ according to the three periods. Thus, a player's performance in various settings can be simulated for illustration. Similarly, Figures 7 and 8 and Figures 9–12 were illustrated by calculating predicted winning rates in each age and period. Figures 9–12 reflect the ageing effect on performance. Figure 9, using the full sample that covers all ages, shows a downward sloping in all the periods. This shows that a player's performance deteriorates over time with a decline in mental strength. Figure 7 demonstrates that winning rates declined as the debut age increased in the three periods. Hence, *Hypothesis 1* is supported, regardless of the environment of the professional Shogi.

The slope in the AI diffusion period is steeper than in any other period. Figure 10 using the sub-sample demonstrates the slope at 45 years because the sample is limited to players who are equal to or below 45 years. In Figure 10, the slope is steeper in the AI diffusion period than other periods, although the winning rates at approximately 20 years are remarkably higher in the AI diffusion period than in other periods. Naturally, in the AI diffusion period, young players are more likely to win than in other periods, whereas older players are less likely to win. This is consistent with *Hypothesis 2*.

Let us see Figures 11 and 12, which are illustrated on the results of Table 4 Panel A and B. In Figure 11, the slope in the AI diffusion period is almost the same as in the period directly before the emergence of AI. Additionally, the level of winning rates does not differ between periods if we compare it at the same age. This means that technology is unlikely to change the ageing effect on the performance of players with low innate abilities. Switching our attention to Figure 12, using the sample of players with high innate abilities, surprisingly, the performance at approximately 20 years is similar between periods. However, the gap drastically increased because the slope is far steeper, especially in the AI diffusion period. Meanwhile, the

---

$0.586$ (=Mean winning rate at 21)+ $\{-0.029(=\alpha_5)\times 5+0.00053(=\alpha_6)\times 5^2\}/2=0.519$
The variables of the second player's debut age are not statistically significant, so its marginal effects are not used in the calculation. Similarly, other predicted winning rates are calculated and used for illustrating Figure 7.

slope in the low-technology period is flatter than in any other period. Figures 11 and 12 jointly imply that the diffusion of AI drastically reduced the advantage of high innate ability as players get older.

The strategy automatically recommended by AI is far better than the strategy that many players think. However, the high ability players' skills and strategies are better than low ability ones. Hence, low innate ability players can enjoy the benefit of AI more than high ability players. Adopting the strategy recommended by AI compensates for the skill gap between the low and innate ability players. This implies that AI is a substitute for the innate ability of the Professional Shogi players, reducing the performance gap between the high and low ability players.

In international chess, Bertoni et al. (2015) made it evident that players' productivity increased rapidly to peak age at 21 years and then declined gradually and constantly. Furthermore, many players dropped out of play before the peak age (Bertoni et al. 2015). Several Shogi players have been forced to drop out before becoming professional players when they belonged to the Shoreikai. Therefore, on average, players debut as professional at 21 years old, equivalent to the peak age of international Chess players. Consistent with the case of international Chess, as illustrated by the simulation in Figures 9–12, the Shogi players winning rates declines constantly after they turn 21 years old. Before entering the professional league, players are thought to improve their skills and performance in the semi-professional Shoreikai league. However, they do not play with professional players, except for some exhibition matches.

The finding of this study that Shogi players' performance declined consistently is in line with the findings of international chess studies that the age-productivity relationship is an inverted U-shape profile, with a peak at approximately 21 years for the case of chess (Bertoni et al. 2015). The relation of FI drivers' age-productivity relationship is also an inverted U-shape. However, its peak is at approximately 31 years (Castellucci et al. 2011). Interestingly, the mental productivity of the Shogi and Chess players declined by the age of 10, younger than that of FI drivers.

## 6. Conclusion

The development and diffusion of AI result in drastic changes in the work environment in the professional Shogi world. In response to this, the professional Shogi players changed their work style to survive as professional players. This study quantitatively examined the impact of technological progress on the performance gaps between players using the games level-data covering the periods before and after the technological progress.

In 2018–2019, twin peaks appeared in the Shogi players' performance proving performance polarisation. In contrast, the performance is distributed in the form of normal distribution if we exclude the sample from 2018 to 2019. We found through regression analysis that (1) AI narrowed the gap of innate ability among same-age players leading to a reduction in the performance gap among them, (2) a consistent decline in the players' winning rate is observed as they got older, (3) the ageing declination of the probability of winning are observed in multiple periods and (4) the AI effects on the ageing declination of the probability of winning is only observed for high innate skill players. These findings suggest that the diffusion of AI caused players to retire from active play earlier in their career, whereas the gap in innate ability to survive as active players was reduced. That is, active players using AI can be considered a substitute for the innate ability of the individual player. These imply that the polarisation of players' performance depends on whether players can make the best use of AI rather than innate ability.

In 1996, Shuji Sato, a middle-class professional Shogi player, confessed, 'I only can pray that AI never wins against professional Shogi players because AI surpassing humans leads demand for professional players to disappear' (Japan Shogi Federation (Nihon Shogi Renmei), 1996). As opposed to his pessimistic view, AI actually made professional Shogi more popular and increased its demand. This is partly because AI numerically decides which player is better in the game. So viewers of the Shogi game can enjoy the game by predicting the winning probability even if they do not have any discernment. AI makes professional Shogi more attractive to everybody and a popular form of entertainment. The professional Shogi game becomes like a professional sports game, such as a professional tennis game. There was a tag-match tournament where tag teams of professional Shogi players and AI participated. This agrees with the discussion that collaboration between humans and AI enhances creativity (Fujita, 2018). AI increases the demand for professional Shogi games. Meanwhile, the professional Shogi players came to retire earlier than ever before. It becomes more difficult to catch up with the speed of change of the newly established strategy if players get older. Naturally, players are more likely to be replaced by younger players. Hence, the mental productivity of the Shogi players is more likely to be physically productive like that of professional sports players.

What we observed is obtained under the novel setting of the labour market of professional Shogi. It is unknown whether the effect of AI diffusion reduced the inequality caused by the innate ability in the settings of other brain works. Future research is needed to analyse how and the extent to which AI diffusion is a substitute for innate ability.

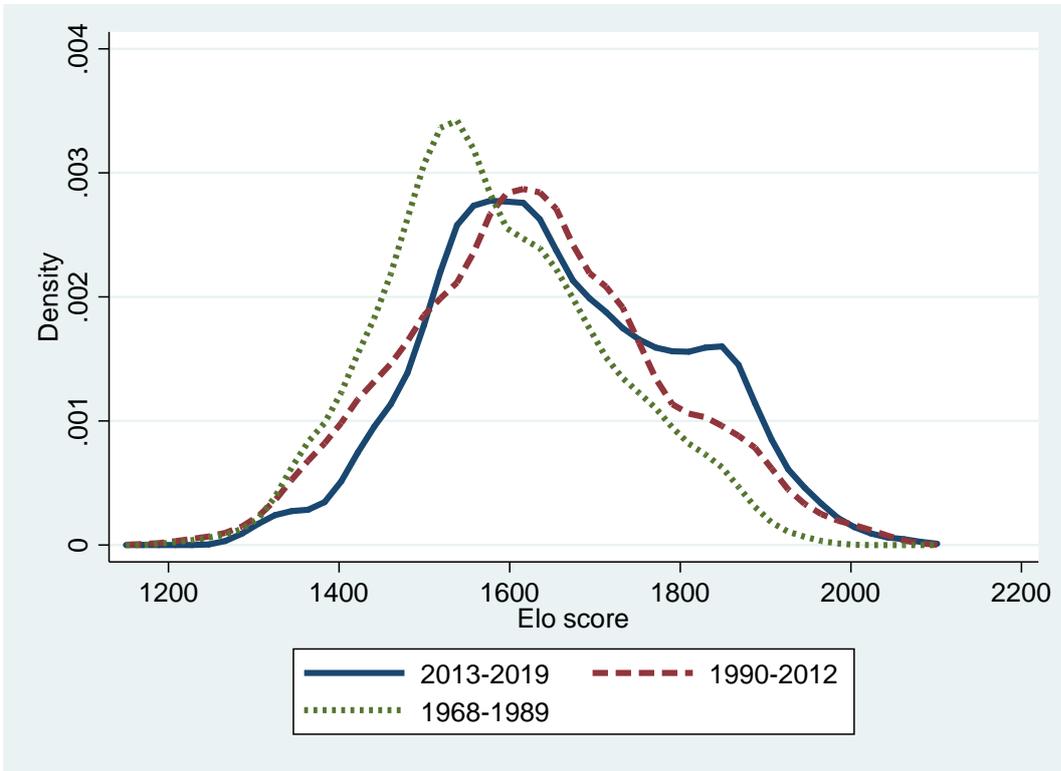

Figure 1. Kernel distribution of performance score (Elo rate): Period 1968–2019

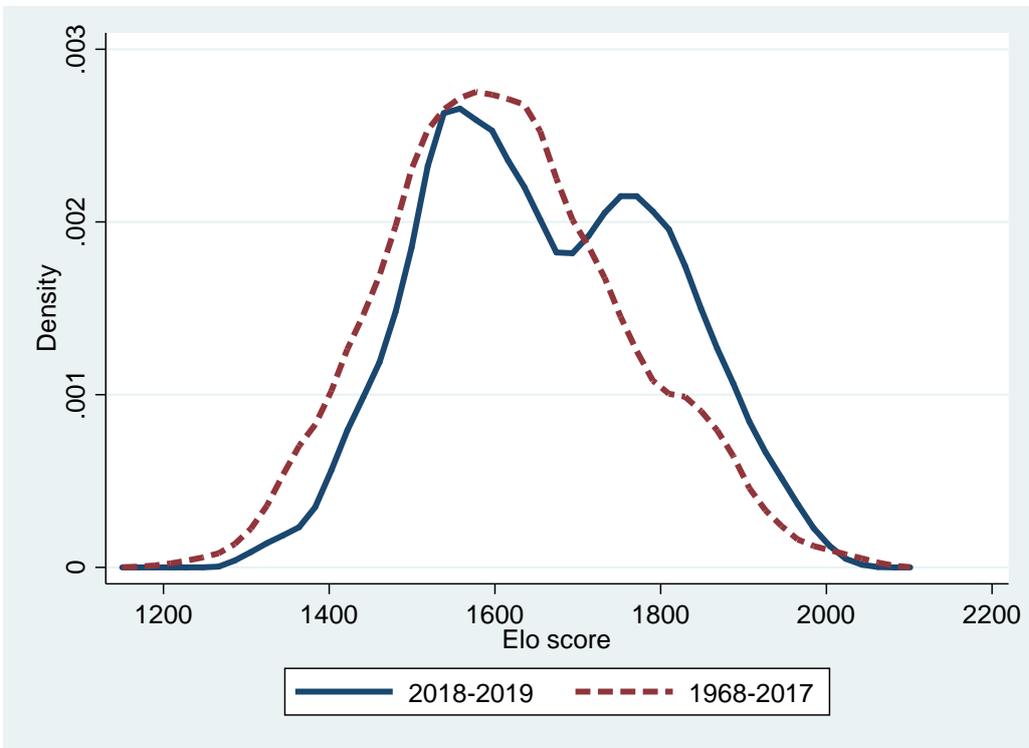

Figure 2. Kernel distribution of performance score (Elo rate): 1968–2017 vs. 2018–2019

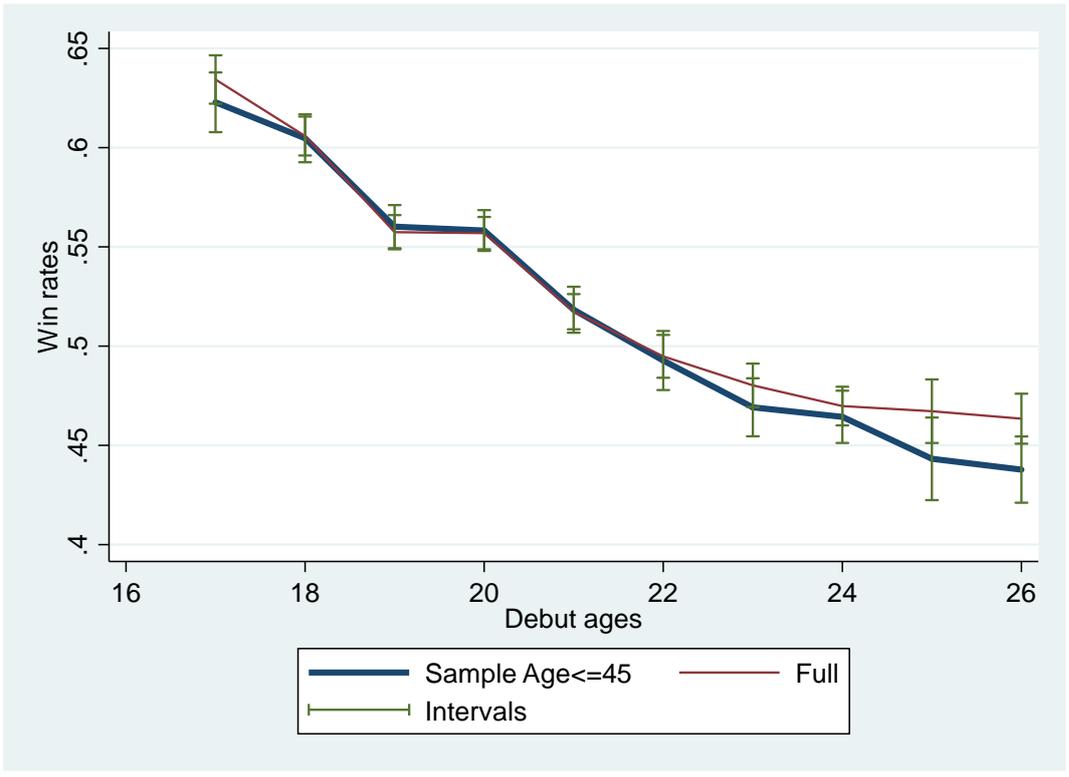

Figure 3. Change in player's mean win rates according to debut ages.

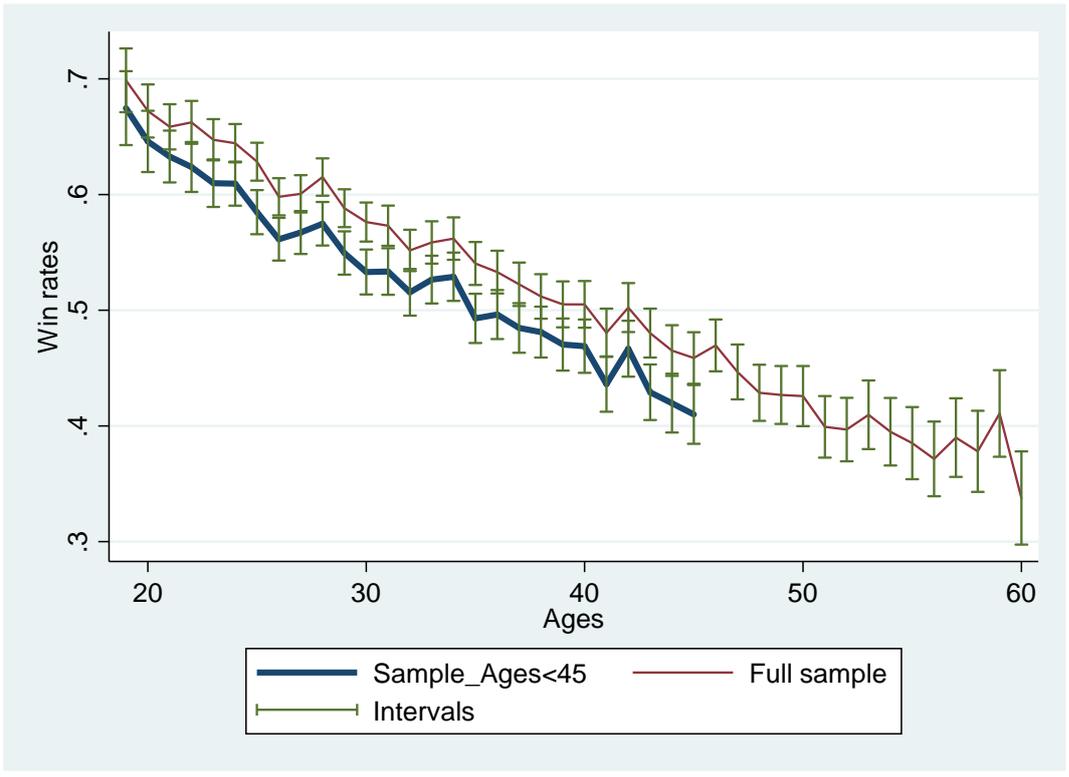

Figure 4. Change in player's mean win rates according to ages.

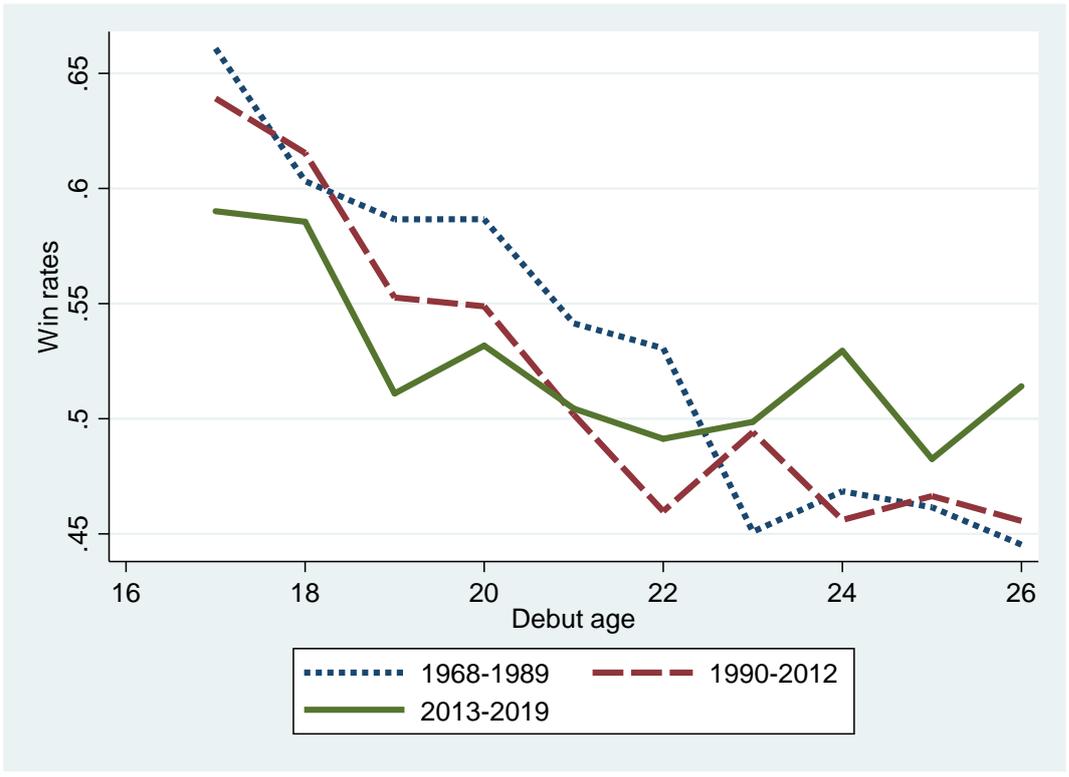

Figure 5. Comparison between the three periods about change in player's mean win rates according to debut ages.

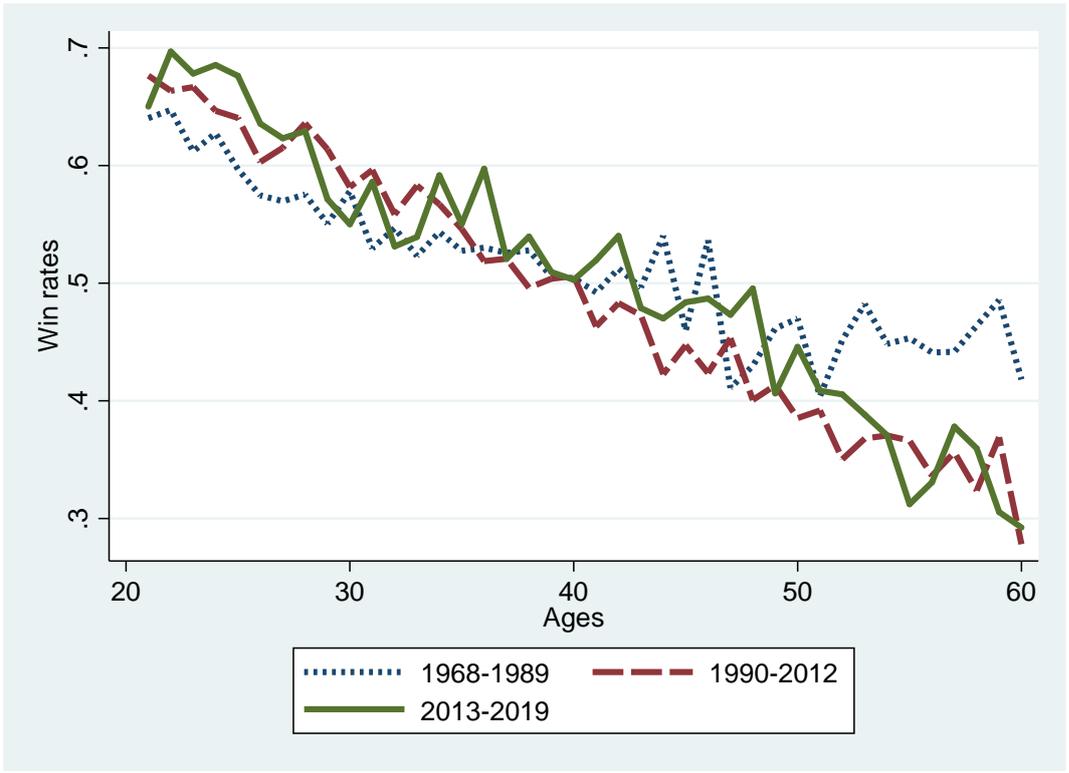

Figure 6. Comparison between the three periods about change in player's mean win rates according to ages.

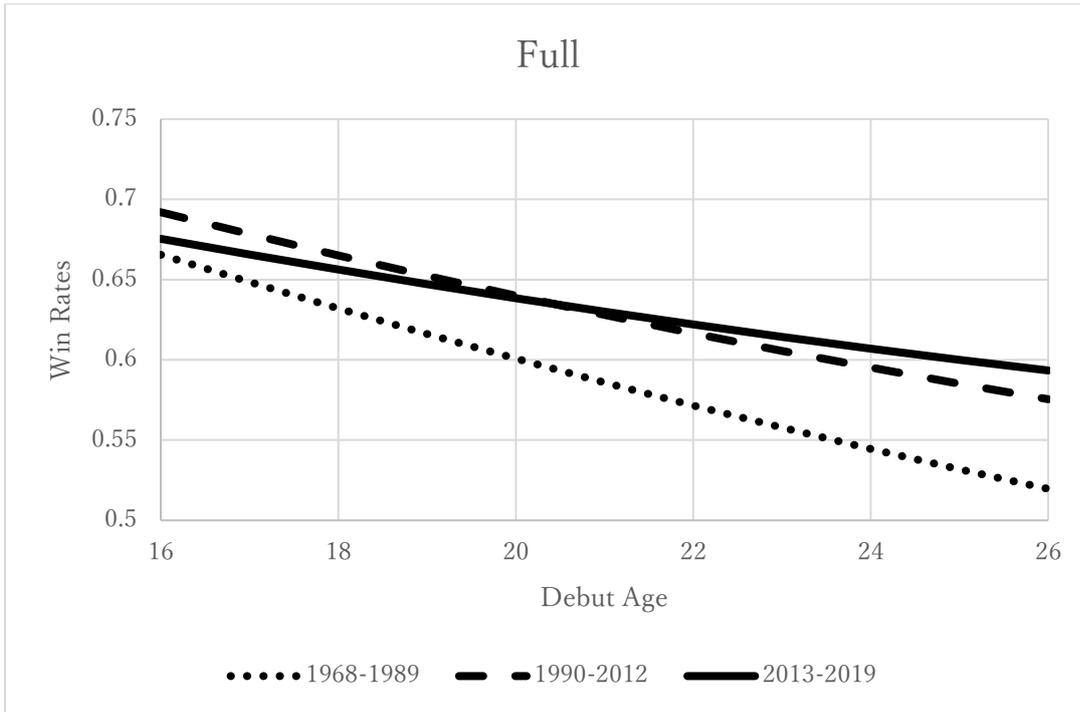

Figure 7. Changes in performance according to debut ages. Sample: All ages.

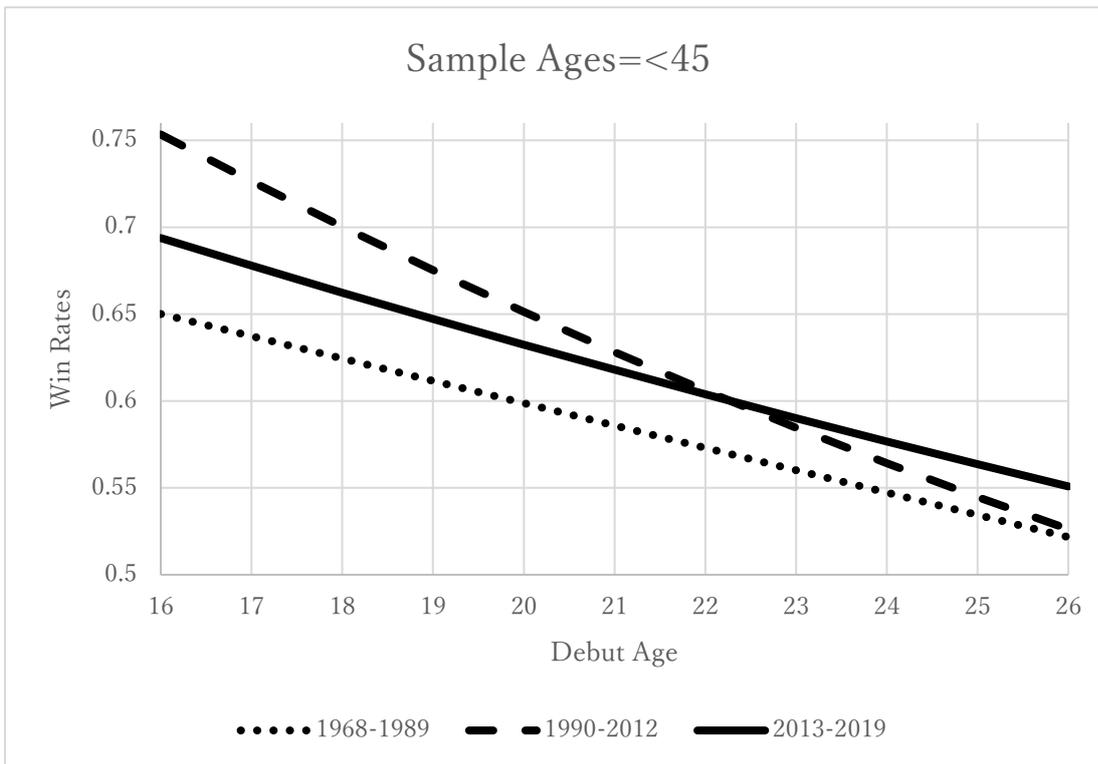

Figure 8. Changes in performance according to debut ages. Sample: Ages<45

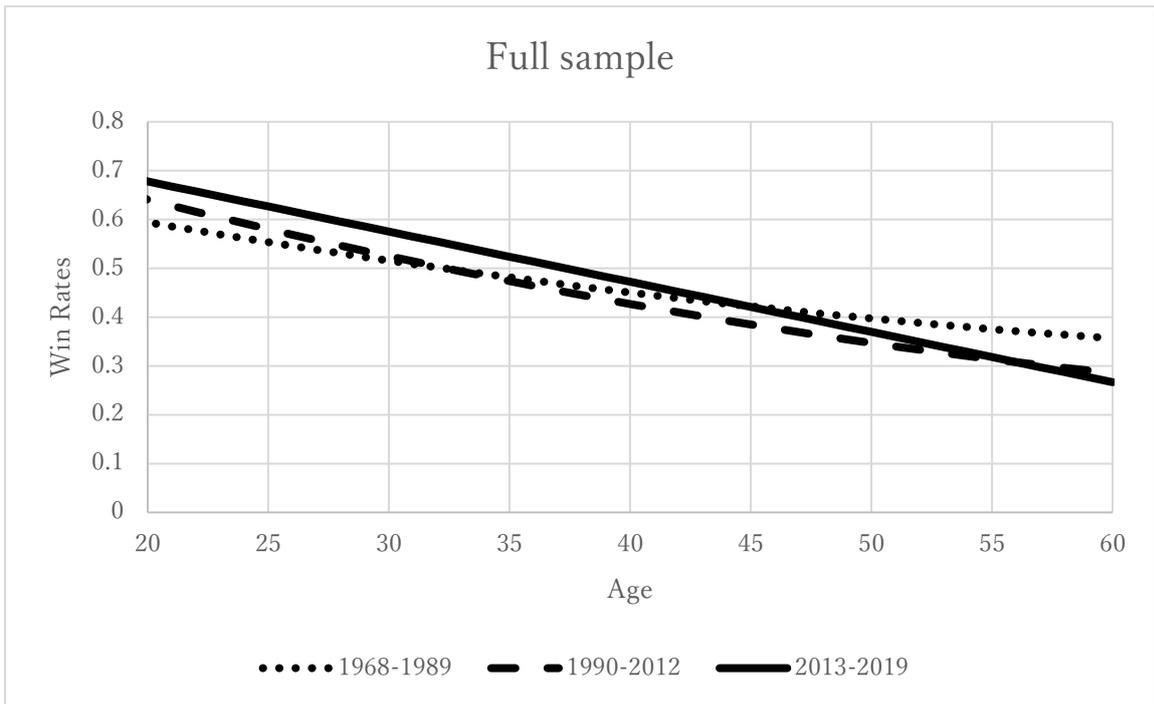

Figure 9 Changes in performance according to ages. Sample: all ages.

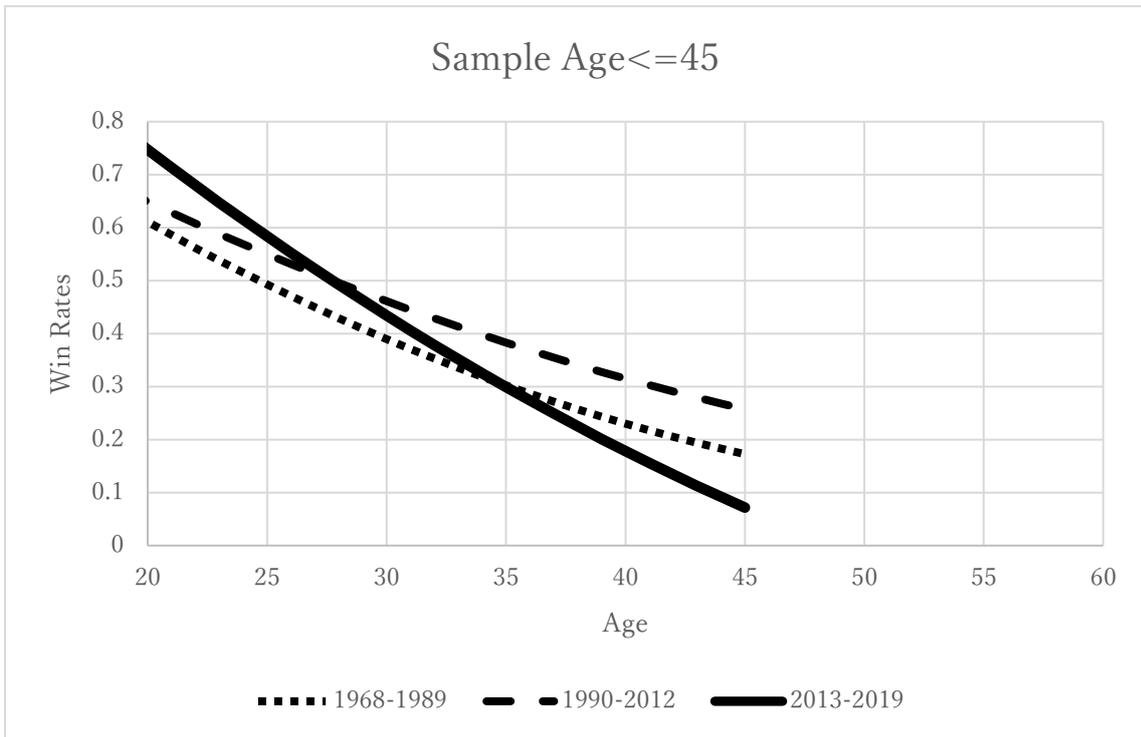

Figure 10 Changes in performance according to ages. Sub-sample: Ages<45

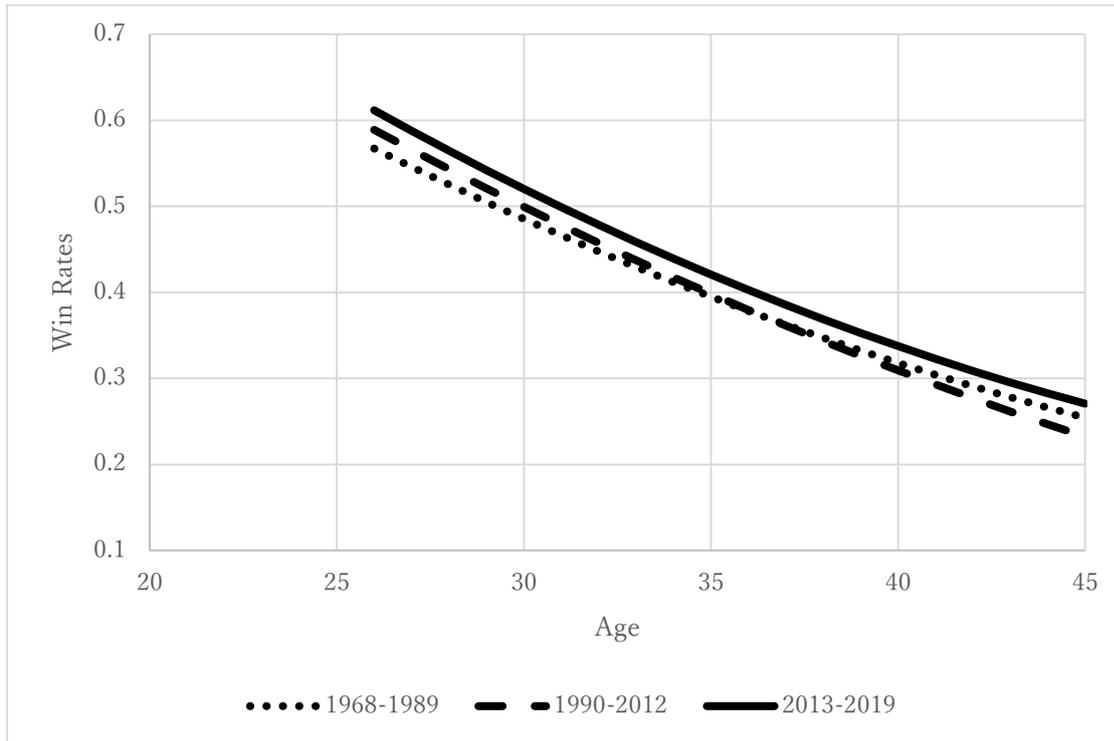

Figure 11. Changes in performance according to age. Sample: Ages<45 and low innate ability

Note: Low innate ability is defined as players who debuted at an age equal to or after 20 years. There are players' records from 26 years at best.

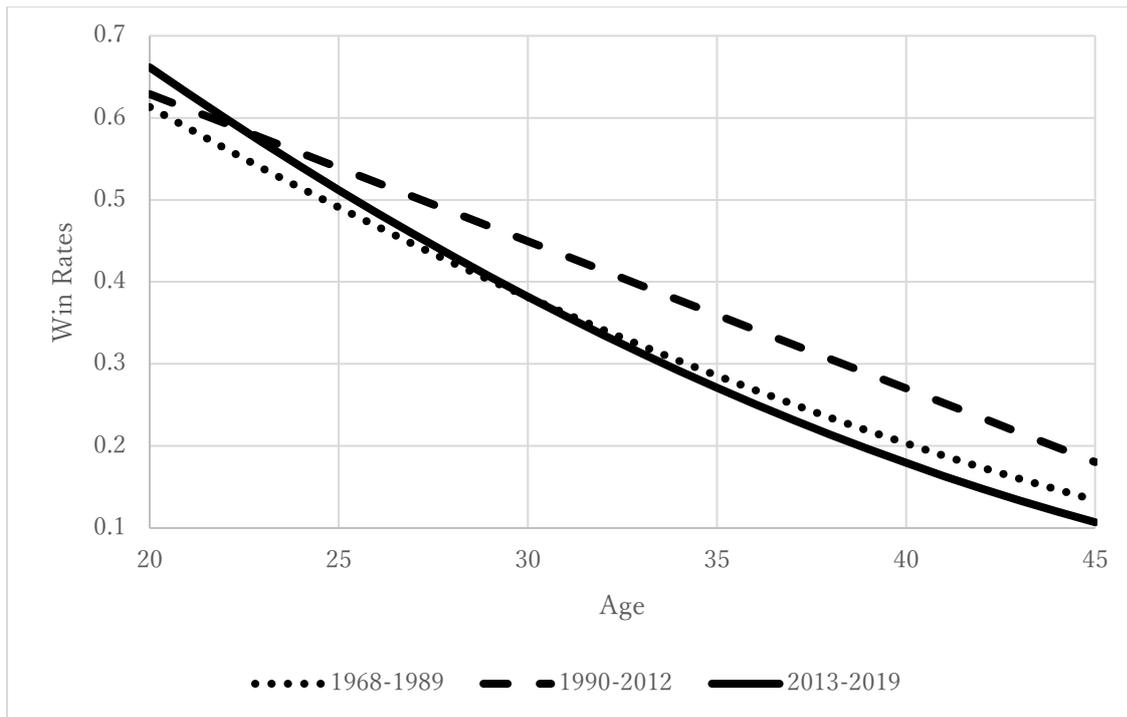

Figure 12. Changes in performance according to age. Sample: Ages<45 and high innate ability
Note: High innate ability is defined as players who debut before or equal to 19.
There are players' records from 20 years at least.

**Table 1: Definitions of key variables and their basic statistics.**

| Variables | Definition | Mean | s.d. | Min. | Max. |
|---|---|---|---|---|---|
| *Dependent Variables* | | | | | |
| WIN | Equals 1 if the first players win, 0 otherwise. | 0.53 | 0.49 | 0 | 1 |
| *Independent Variables* | | | | | |
| AGE | The first players' ages | 36.7 | 11.5 | 15 | 76 |
| $AGE^2/100$ | Square of the first players' ages/ 100 | --- | --- | --- | --- |
| AGE_OP | The second (opponent) players' ages | 37.8 | 11.8 | 15 | 77 |
| $AGE\_OP^2/100$ | Square of the second (opponent) players' ages/ 100 | --- | --- | --- | --- |
| DEB_AGE | The first players' debut ages | 20.9 | 2.65 | 15 | 26 |
| $DEB\_AGE^2/100$ | Square of the first players' debut ages/ 100 | --- | --- | --- | --- |
| DEB_AGE_OP | The second (opponent) players' debut ages | 21.3 | 3.52 | 15 | 26 |
| $DEB\_AGE\_OP^2/100$ | Square of the second (opponent) players' debut ages/ 100 | --- | --- | --- | --- |
| ELO/100 | The first players' Elo Score/100 | 16.2 | 1.41 | 11.6 | 20.9 |
| ELO_OP/100 | The second (opponent) players' Elo Score | 16.2 | 1.45 | 11.7 | 20. |

Note: Draw games are excluded from the sample. Sample covers ages over 45 and its observations are 88,788

Table 2. **Dependent variable: WIN (Probit model): Sample covered all ages.**

|  | (1) *1968-1989* | (2) *1990-2012* | (3) *2013-2019* |
|---|---|---|---|
| AGE | −0.008*** | −0.010*** | −0.011*** |
|  | (0.002) | (0.002) | (0.003) |
| AGE$^2$/100 | 0.006** | 0.006*** | 0.006 |
|  | (0.002) | (0.002) | (0.004) |
| AGE_OP | 0.008*** | 0.014*** | 0.009*** |
|  | (0.002) | (0.002) | (0.003) |
| AGE_OP$^2$/100 | −0.007*** | −0.012*** | −0.006 |
|  | (0.002) | (0.002) | (0.004) |
| DEB_AGE | −0.029*** | −0.023*** | −0.016 |
|  | (0.006) | (0.007) | (0.013) |
| DEB_AGE$^2$/100 | 0.053*** | 0.045*** | 0.039 |
|  | (0.011) | (0.002) | (0.028) |
| DEB_AGE_OP | 0.003 | 0.010 | 0.018* |
|  | (0.006) | (0.008) | (0.010) |
| DEB_AGE_OP$^2$/100 | 0.003 | −0.018 | −0.034* |
|  | (0.010) | (0.019) | (0.021) |
| ELO/100 | 0.097*** | 0.102*** | 0.091*** |
|  | (0.004) | (0.004) | (0.007) |
| ELO_OP/100 | −0.100*** | −0.094*** | −0.099*** |
|  | (0.004) | (0.004) | (0.007) |
| Pseudo R$^2$ | 0.08 | 0.10 | 0.10 |
| Observations | 33,212 | 43,077 | 12,499 |

Note: ***, ** and * denote statistical significance at the 1%, 5%, and 10% levels, respectively. Numbers without parentheses are marginal effects. Various control variables are included in all columns, such as dummies for a player's and his opponent's rank from 4 dan to 9 dan, the game's status dummies, and the year when the game was held. In addition, game status dummies are made based on the following tournaments or leagues: 8 major title-match leagues (or tournament) (1) Ryuo, (2) Meijin, (3) Oi, (4) Oza, (5) Kio, (6) Eio, (7) Osho, (8) Kisei. Non-major title tournaments such (9) Asahi-hai, (10) NHK-hai, (11) Ginga-sen, (12) Japan Professional-hai (13) Shinjin-o, (13) YAMADA challenge-hai, (14) Kakogawa-sen. However, these estimates are not reported.

Table 3. **Dependent variable: WIN (Probit model): Sub-sample of players below 45 ages.**

|  | (1) *1968-1989* | (2) *1990-2012* | (3) *2013-2019* |
|---|---|---|---|
| AGE | −0.023*** | −0.022*** | −0.022** |
|  | (0.006) | (0.005) | (0.009) |
| AGE$^2$/100 | 0.031*** | 0.024*** | 0.022 |
|  | (0.009) | (0.007) | (0.014) |
| AGE_OP | 0.026*** | 0.019*** | 0.043*** |
|  | (0.006) | (0.005) | (0.009) |
| AGE_OP$^2$/100 | −0.030*** | −0.018** | −0.057*** |
|  | (0.009) | (0.008) | (0.012) |
| DEB_AGE | −0.026* | −0.024** | −0.013 |
|  | (0.014) | (0.011) | (0.011) |
| DEB_AGE$^2$/100 | 0.045 | 0.051** | 0.035* |
|  | (0.032) | (0.023) | (0.019) |
| DEB_AGE_OP | −0.007 | 0.021* | 0.030* |
|  | (0.012) | (0.011) | (0.018) |
| DEB_AGE_OP$^2$/100 | 0.018 | −0.044* | −0.057 |
|  | (0.029) | (0.023) | (0.037) |
| ELO/100 | 0.097*** | 0.104*** | 0.099*** |
|  | (0.007) | (0.004) | (0.009) |
| ELO_OP/100 | −0.094*** | −0.092*** | −0.084*** |
|  | (0.006) | (0.005) | (0.009) |
| Pseudo R$^2$ | 0.07 | 0.08 | 0.07 |
| Observations | 17,881 | 27,933 | 6,416 |

Note: Note: ***, ** and * denote statistical significance at the 1%, 5%, and 10% levels, respectively. Numbers without parentheses are marginal effects. In all columns, control variables included in estimations of Table 2 are included, although these estimates are not reported.

Table 4. **Dependent variable: WIN (Probit model): Sub-sample of players below 45**

Panel A. Low innate ability (DEB_AGE<=20)

|  | (1) *1968-1989* | (2) *1990-2012* | (3) *2013-2019* |
|---|---|---|---|
| AGE | − 0.019** | − 0.029*** | − 0.018 |
|  | (0.009) | (0.007) | (0.016) |
| AGE$^2$/100 | 0.023* | 0.033*** | 0.015 |
|  | (0.014) | (0.010) | (0.024) |
| AGE_OP | 0.024*** | 0.018*** | 0.048*** |
|  | (0.007) | (0.006) | (0.013) |
| AGE OP$^2$/100 | − 0.026** | − 0.015* | − 0.065*** |
|  | (0.011) | (0.008) | (0.018) |
| DEB_AGE | −0.151*** | −0.010 | −0.029 |
|  | (0.039) | (0.019) | (0.019) |
| DEB_AGE$^2$/100 | 0.296** | 0.021 | 0.065* |
|  | (0.077) | (0.039) | (0.038) |
| DEB_AGE_OP | 0.010 | 0.006 | 0.023 |
|  | (0.016) | (0.015) | (0.021) |
| DEB_AGE_OP$^2$/100 | −0.023 | −0.011 | −0.045 |
|  | (0.035) | (0.033) | (0.043) |
| ELO/100 | 0.091*** | 0.108*** | 0.114*** |
|  | (0.009) | (0.005) | (0.012) |
| ELO_OP/100 | −0.095*** | −0.093*** | −0.806*** |
|  | (0.007) | (0.006) | (0.012) |
| Pseudo R$^2$ | 0.07 | 0.08 | 0.07 |
| Observations | 13,195 | 16,398 | 3,830 |

Panel B. Sample: High innate ability (DEB_AGE>20)

|  | (1) *1968-1989* | (2) *1990-2012* | (3) *2013-2019* |
|---|---|---|---|
| AGE | −0.028** | −0.015** | −0.033** |
|  | (0.011) | (0.007) | (0.014) |
| AGE$^2$/100 | 0.037** | 0.014 | 0.038* |
|  | (0.017) | (0.009) | (0.021) |
| AGE_OP | 0.024** | 0.021** | 0.030** |
|  | (0.011) | (0.009) | (0.012) |
| AGE OP$^2$/100 | −0.032** | −0.023 | −0.039** |
|  | (0.016) | (0.014) | (0.016) |
| DEB_AGE | −0.025 | −0.043 | −0.318* |
|  | (0.101) | (0.078) | (0.172) |
| DEB_AGE$^2$/100 | 0.070 | 0.092 | 0.918* |
|  | (0.327) | (0.237) | (0.496) |
| DEB_AGE_OP | −0.050*** | 0.041*** | 0.049 |
|  | (0.018) | (0.014) | (0.034) |
| DEB_AGE_OP$^2$/100 | 0.127*** | −0.089*** | −0.098 |
|  | (0.044) | (0.033) | (0.075) |
| ELO/100 | 0.091*** | 0.093*** | 0.078*** |
|  | (0.011) | (0.006) | (0.014) |
| ELO_OP/100 | −0.081*** | −0.088*** | −0.083*** |
|  | (0.012) | (0.006) | (0.015) |
| Pseudo R$^2$ | 0.06 | 0.07 | 0.06 |
| Observations | 4,686 | 10,995 | 2,586 |

Note: Note: ***, ** and * denote statistical significance at the 1%, 5%, and 10% levels, respectively. Numbers without parentheses are marginal effects. In all columns, control variables included in estimations of Table 2 are included, although these estimates are not reported.

Appendix

Figure A1. Change in player's productivity according to ages.

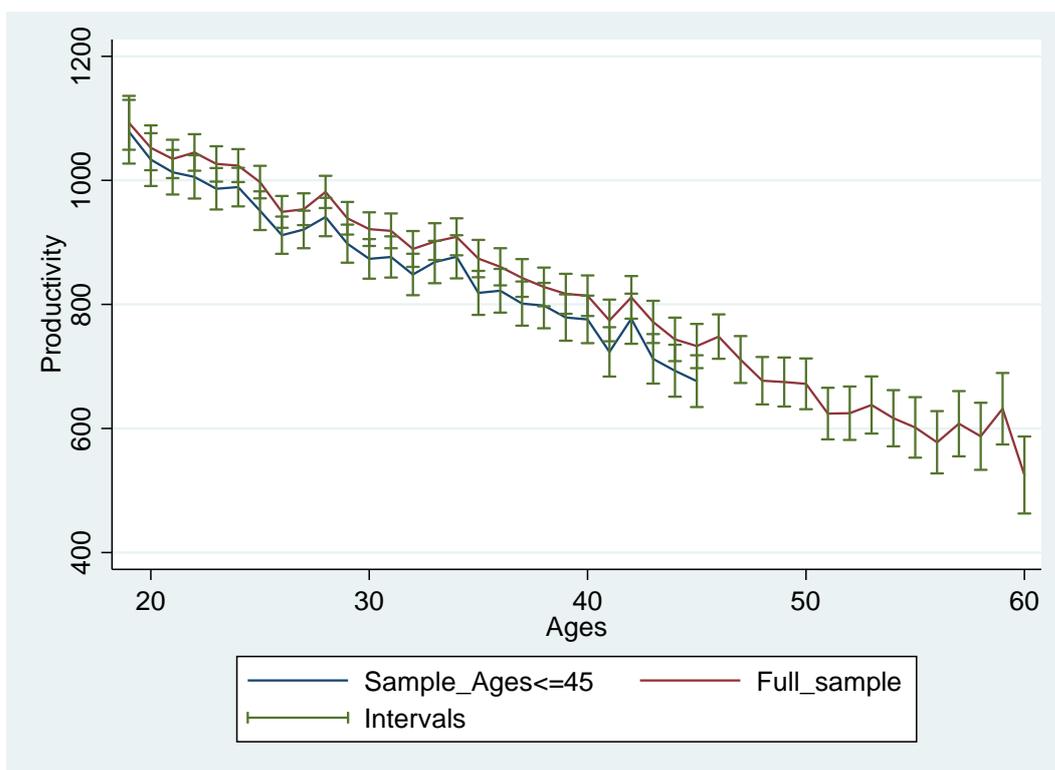

Note: The productivity index is calculated in the way of Bertoni et al. (2015, p.48).